\documentclass[12pt,a4paper]{article}
\makeatletter
\def\eqnarray{%
\stepcounter{equation}%
\let\@currentlabel=\theequation
\global\@eqnswtrue
\global\@eqcnt\z@
\tabskip\@centering
\let\\=\@eqncr
$$\halign to \displaywidth\bgroup\@eqnsel\hskip\@centering
$\displaystyle\tabskip\z@{##}$&\global\@eqcnt\@ne
\hfil$\displaystyle{{}##{}}$\hfil
&\global\@eqcnt\tw@$\displaystyle\tabskip\z@{##}$\hfil
\tabskip\@centering&\llap{##}\tabskip\z@\cr}
\makeatother
\newcommand{\fukuso}{{\mathbf C}}
\newcommand{\futon}{{\bf N}}

\begin{document}

\title{\sl A Modern Introduction to Cardano and Ferrari Formulas 
in the Algebraic Equations}
\author{
  Kazuyuki FUJII 
  \thanks{E-mail address : fujii@yokohama-cu.ac.jp }\\
  Department of Mathematical Sciences\\
  Yokohama City University\\
  Yokohama, 236--0027\\
  Japan
  }
\date{}
\maketitle
%
%
%
%
\begin{abstract}
  We give a modern approach to the famous Cardano and Ferrari formulas 
  in the algebraic equations with three and four degrees. Namely, 
  we reconstruct these formulas from the point of view of superposition 
  principle in quantum computation based on three and four level systems 
  which are being developed by the author. 
  
  We also present a problem on some relation between Galois theory and 
  Qudit theory. 
\end{abstract}

\newpage

%
%
%
%
\section{Introduction}

The method to obtain algebraic solutions in (algebraic) equations has a very 
long history, see \cite{MSJ}. 
To solve a quadratic equation is very easy. 
To solve a cubic equation is not so easy and has been given by Cardano. 
How to solve a quartic equation is comparatively hard and Ferrari has given 
such solutions. 
After that many mathematicians including Lagrange, Gauss et al made 
how to solve equations refined. 
However, (algebraic) equations with degrees of more than four have in general 
no algebraic solutions, which was proved by two young mathematicians Abel and 
Galois. 

By the way, we are studying a quantum computation based on multi--level 
systems of (laser--cooled) atoms. The three and four level systems are 
particularly interesting. In the theories the generalized Pauli and 
Walsh--Hadamard matrices (gates) play a central role. We don't mention 
the topics, see for example \cite{KF1}--\cite{KuF} as an introduction. 

The aim of this note is to show that some fundamental techniques of 
the generalized Pauli and Walsh--Hadamard matrices are applicable to 
construct the Caldano and Ferrari formulas. The essential point is 
{\bf the principle of superposition} which is the heart of Quantum 
Computation. 
Namely, we reconstruct the Caldano and Ferrari formulas from the view point 
of this principle. 

Our method is so simple that it must have been known in some context. 
However, we cannot check all related references because the history is too 
long! We could not find such references within our effort. 
Anyway, by this note many students will be able to master the Caldano and 
Ferrari formulas without much effort.

\section{Mathematical Preliminaries}
We summarize the properties of Pauli matrices and Walsh--Hadamard matrix, 
and next state the corresponding ones of generalized Pauli matrices and 
Walsh--Hadamard matrix within our necessity. 

Let $\{\sigma_{1}, \sigma_{2}, \sigma_{3}\}$ be Pauli matrices : 
\begin{equation}
\label{eq:pauli}
\sigma_{1} = 
\left(
  \begin{array}{cc}
    0& 1 \\
    1& 0
  \end{array}
\right), \quad 
\sigma_{2} = 
\left(
  \begin{array}{cc}
    0& -i \\
    i& 0
  \end{array}
\right), \quad 
\sigma_{3} = 
\left(
  \begin{array}{cc}
    1& 0 \\
    0& -1
  \end{array}
\right).
\end{equation}
By (\ref{eq:pauli}) $\sigma_{2}=i\sigma_{1}\sigma_{3}$, so that the essential 
elements of Pauli matrices are $\{\sigma_{1}, \sigma_{3}\}$ and they satisfy
\begin{equation}
\sigma_{1}^{2}=\sigma_{3}^{2}={\bf 1}_{2}\ ;\quad 
\sigma_{1}^{\dagger}=\sigma_{1},\
\sigma_{3}^{\dagger}=\sigma_{3}\ ;\quad 
\sigma_{3}\sigma_{1}=-\sigma_{1}\sigma_{3}=\mbox{e}^{i\pi}\sigma_{1}\sigma_{3}.
\end{equation}

The Walsh--Hadamard matrix is defined by 
\begin{equation}
   \label{eq:w-a}
   W = \frac{1}{\sqrt{2}}
     \left(
        \begin{array}{rr}
            1& 1 \\
            1& -1
        \end{array}
     \right)\ \in \ O(2)\ \subset U(2).
\end{equation}
This matrix (or transformation) is unitary and it plays a very important role 
in Quantum Computation. Moreover it is easy to realize it in Quantum Optics 
as shown in \cite{KF4}. 
Let us list some important properties of $W$ :
\begin{eqnarray}
\label{eq:properties of W-H (1)}
      &&W^{2}={\bf 1}_{2},\ \ W^{\dagger}=W=W^{-1}, \\
\label{eq:properties of W-H (2)}
      &&\sigma_{1}= W\sigma_{3}W^{-1},
\end{eqnarray}
The check is very easy. 

Next let us generalize Pauli matrices to higher dimensional cases. 
Let $\{\Sigma_{1}, \Sigma_{3}\}$ be the following matrices in $M(n,\fukuso)$

\begin{equation}
\label{eq:gener-pauli}
\Sigma_{1}=
\left(
\begin{array}{cccccc}
0&  &  &      &      &       1   \\
1& 0&  &      &      &           \\
  & 1& 0&      &      &          \\
  &  & 1& \cdot&      &          \\
  &  &  & \cdot& \cdot&          \\
  &  &  &      &    1 & 0
\end{array}
\right),      \qquad
\Sigma_{3}=
\left(
\begin{array}{cccccc}  
1&        &           &      &      &                  \\
  & \sigma&           &      &      &                  \\
  &       & {\sigma}^2&      &      &                  \\
  &       &           & \cdot&      &                  \\
  &       &           &      & \cdot&                  \\
  &       &           &      &      &  {\sigma}^{n-1}
\end{array}
\right)
\end{equation}
where $\sigma$ is a primitive root of unity ${\sigma}^{n}=1$ (
$\sigma=\mbox{e}^{\frac{2\pi i}{n}}$). We note that
\[
\bar{\sigma}=\sigma^{n-1},\quad
1+\sigma+\cdots+\sigma^{n-1}=0 .
\]
The two matrices
$\{\Sigma_{1}, \Sigma_{3}\}$ are generalizations of Pauli matrices
$\{\sigma_{1}, \sigma_{3}\}$, but they are not hermitian.
Here we list some of their important properties :
\begin{equation}
\Sigma_{1}^{n}=\Sigma_{3}^{n}={\bf 1}_{n}\ ; \quad
\Sigma_{1}^{\dagger}=\Sigma_{1}^{n-1},\
\Sigma_{3}^{\dagger}=\Sigma_{3}^{n-1}\ ; \quad
\Sigma_{3}\Sigma_{1}=\sigma \Sigma_{1}\Sigma_{3}\ .
\end{equation}
For $n=3$ and $n=4$ $\Sigma_{1}$ and its powers are given respectively as 
\begin{equation}
\label{eq:sigma-1-three}
\Sigma_{1}=
\left(
\begin{array}{ccc}
     0 &   & 1   \\
     1 & 0 &     \\
       & 1 & 0
\end{array}
\right),\quad 
\Sigma_{1}^{2}=
\left(
\begin{array}{ccc}
     0 & 1 &     \\
       & 0 & 1   \\
     1 &   & 0
\end{array}
\right)       
\end{equation}
and

\begin{equation}
\label{eq:sigma-1-four}
\Sigma_{1}=
\left(
\begin{array}{cccc}
   0 &   &    & 1  \\
   1 & 0 &    &    \\
     & 1 & 0  &    \\
     &   & 1  & 0
\end{array}
\right),\quad 
\Sigma_{1}^{2}=
\left(
\begin{array}{cccc}
   0 &   & 1  &    \\
     & 0 &    & 1  \\
   1 &   & 0  &    \\
     & 1 &    & 0
\end{array}
\right),\quad 
\Sigma_{1}^{3}=
\left(
\begin{array}{cccc}
   0 & 1 &    &    \\
     & 0 & 1  &    \\
     &   & 0  & 1  \\
   1 &   &    & 0
\end{array}
\right).
\end{equation}

If we define a Vandermonde matrix $W$ based on $\sigma$ as
\begin{eqnarray}
\label{eq:Large-double}
W&=&\frac{1}{\sqrt{n}}
\left(
\begin{array}{ccccccc}
1&        1&     1&   \cdot & \cdot  & \cdot & 1             \\
1& \sigma^{n-1}& \sigma^{2(n-1)}&  \cdot& \cdot& \cdot& \sigma^{(n-1)^2} \\
1& \sigma^{n-2}& \sigma^{2(n-2)}&  \cdot& \cdot& \cdot& \sigma^{(n-1)(n-2)} \\
\cdot&  \cdot &  \cdot  &     &      &      & \cdot  \\
\cdot&  \cdot  & \cdot &      &      &      &  \cdot  \\
1& \sigma^{2}& \sigma^{4}& \cdot & \cdot & \cdot & \sigma^{2(n-1)} \\
1& \sigma & \sigma^{2}& \cdot& \cdot& \cdot& \sigma^{n-1}
\end{array}
\right), \\
\label{eq:Large-double-dagger}
W^{\dagger}&=&\frac{1}{\sqrt{n}}
\left(
\begin{array}{ccccccc}
1&        1&     1&   \cdot & \cdot  & \cdot & 1             \\
1& \sigma& \sigma^{2}&  \cdot& \cdot& \cdot& \sigma^{n-1} \\
1& \sigma^{2}& \sigma^{4}&  \cdot& \cdot& \cdot& \sigma^{2(n-1)} \\
\cdot&  \cdot &  \cdot  &     &      &      & \cdot  \\
\cdot&  \cdot  & \cdot &      &      &      &  \cdot  \\
1& \sigma^{n-2}& \sigma^{2(n-2)}& \cdot& \cdot& \cdot& \sigma^{(n-1)(n-2)} \\
1&    \sigma^{n-1} & \sigma^{2(n-1)}& \cdot& \cdot& \cdot& \sigma^{(n-1)^2}
\end{array}
\right),
\end{eqnarray}
then it is not difficult to see
\begin{equation}
W^{\dagger}W=WW^{\dagger}={\bf 1}_{n},\quad 
\Sigma_{1}=W\Sigma_{3}W^{\dagger}=W\Sigma_{3}W^{-1}.
\end{equation}

Since $W$ corresponds to the Walsh--Hadamard matrix (\ref{eq:w-a}), 
so it may be possible to call $W$ the generalized Walsh--Hadamard matrix. 
If we write $W^{\dagger}=(w_{ab})$, then 
\[
w_{ab}=\frac{1}{\sqrt{n}}\sigma^{ab}=
\frac{1}{\sqrt{n}}\mbox{exp}\left(\frac{2\pi i}{n}ab\right) 
\quad \mbox{for}\quad 0\leq a,\ b \leq n-1. 
\]
This is just the coefficient matrix of Discrete Fourier Transform (DFT) 
if $n=2^{k}$ for some $k \in \futon$, see \cite{PWS}. 

For $n=3$ and $n=4$ $W$ is given respectively as 
\begin{equation}
\label{eq:w-a-three}
W=
\frac{1}{\sqrt{3}}
\left(
\begin{array}{ccc}
1&   1&   1              \\
1& \sigma^{2} & \sigma   \\
1& \sigma & \sigma^{2}
\end{array}
\right)
=
\frac{1}{\sqrt{3}}
\left(
\begin{array}{ccc}
1&          1&                     1                 \\
1& \frac{-1-i\sqrt{3}}{2} & \frac{-1+i\sqrt{3}}{2}   \\
1& \frac{-1+i\sqrt{3}}{2} & \frac{-1-i\sqrt{3}}{2}
\end{array}
\right)
\end{equation}
and 
\begin{equation}
\label{eq:w-a-four}
W=
\frac{1}{2}
\left(
\begin{array}{cccc}
 1 &  1 &  1 &  1                       \\
 1 & \sigma^{3} & \sigma^{2} & \sigma   \\
 1 & \sigma^{2} & 1 & \sigma^{2}        \\
 1 & \sigma & \sigma^{2} & \sigma^{3}
\end{array}
\right)
=
\frac{1}{2}
\left(
\begin{array}{cccc}
 1 &  1 &  1 &  1   \\
 1 & -i & -1 &  i   \\
 1 & -1 &  1 & -1   \\
 1 &  i & -1 & -i
\end{array}
\right).
\end{equation}

\vspace{5mm}
We note that the generalized Pauli and Walsh--Hadamard matrices in three and 
four level systems can be constructed (by using Rabi oscillations of several 
types) in a quantum optical manner, see \cite{KF4} and \cite{KF5}.

\section{Cardano Formula}

We consider an algebraic equation with three degrees 
\begin{equation}
\label{eq:3-degree equation}
x^{3}+ax^{2}+bx+c=0.
\end{equation}
Then by replacing $x\longrightarrow x-(a/3)$ we have 
\[
x^{3}+px+q=0
\]
with $p=b-\frac{a^{2}}{3}$ and $q=c-\frac{ab}{3}+\frac{2a^{3}}{27}$. However, 
we usually set 
\begin{equation}
\label{eq:3-degree equation modify}
x^{3}+3px+q=0
\end{equation}
with 
\[
p=\frac{b}{3}-\frac{a^{2}}{9},\quad q=c-\frac{ab}{3}+\frac{2a^{3}}{27}. 
\]
For this equation we have the famous solutions called {\bf Cardano formula} : 
Three solutions are given by 
\begin{equation}
\label{eq:Cardano formula}
x=u_{0}+v_{0},\quad 
\sigma u_{0}+\sigma^{2}v_{0},\quad 
\sigma^{2}u_{0}+\sigma v_{0},
\end{equation}
where $u_{0}$ and $v_{0}$ are primitive solutions of binomial equations 
\[
u^{3}=\frac{-q+\sqrt{q^{2}+4p^{3}}}{2},\quad 
v^{3}=\frac{-q-\sqrt{q^{2}+4p^{3}}}{2}
\]
and $\sigma=\mbox{e}^{\frac{2\pi i}{3}}$.

\vspace{5mm}
The solutions (\ref{eq:Cardano formula}) look like (classical) superpositions 
of $u_{0}$ and $v_{0}$. From here we reconstruct the solutions by using 
generalized Pauli and Walsh--Hadamard matrices  
(\ref{eq:sigma-1-three}), (\ref{eq:w-a-three}) in three level systems, 
which makes the principle of superposition more clearly. 

Let $x_{1}$, $x_{2}$, $x_{3}$ be three solutions of 
(\ref{eq:3-degree equation modify}) and set 
\begin{equation}
A=
\left(
\begin{array}{ccc}
x_{1} &       &        \\
      & x_{2} &        \\
      &       & x_{3}
\end{array}
\right),\quad x_{j}^{3}+3px_{j}+q=0.
\end{equation}
Remind that $x_{1}+x_{2}+x_{3}=0$ from (\ref{eq:3-degree equation modify}), 
so $\mbox{tr}A=0$. Then we have 
\begin{equation}
\label{eq:3-degree matrix equation}
A^{3}+3pA+qE=0,
\end{equation}
where $E$ is a unit matrix. Now we consider superpositions of solutions 
$x_{j}$ ($j=1, 2, 3$). For that we operate $W$ in (\ref{eq:w-a-three}) on 
(\ref{eq:3-degree matrix equation}) as the adjoint action ($W^{\dagger}W=E$) 
\begin{equation}
\label{eq:3-degree matrix equation adjoint}
(W^{\dagger}AW)^{3}+3p(W^{\dagger}AW)+qE=0.
\end{equation}
Let us calculate $W^{\dagger}AW$ : 
\begin{eqnarray}
\label{eq:calculation-three}
W^{\dagger}AW
&=&
\frac{1}{3}
\left(
\begin{array}{ccc}
 x_{1}+x_{2}+x_{3} & x_{1}+\sigma^{2}x_{2}+\sigma x_{3} & 
 x_{1}+\sigma x_{2}+\sigma^{2}x_{3}                                      \\
 x_{1}+\sigma x_{2}+\sigma^{2}x_{3} & x_{1}+x_{2}+x_{3} & 
 x_{1}+\sigma^{2}x_{2}+\sigma x_{3}                                      \\
 x_{1}+\sigma^{2}x_{2}+\sigma x_{3} & x_{1}+\sigma x_{2}+\sigma^{2}x_{3} 
 & x_{1}+x_{2}+x_{3}
\end{array}
\right)              \nonumber \\
&\equiv&
\left(
\begin{array}{ccc}
     0   &  \beta & \alpha   \\
  \alpha &  0     & \beta    \\
  \beta  & \alpha &   0
\end{array}
\right)
=
\alpha
\left(
\begin{array}{ccc}
     0 &   & 1   \\
     1 & 0 &     \\
       & 1 & 0
\end{array}
\right)+
\beta
\left(
\begin{array}{ccc}
     0 & 1 &     \\
       & 0 & 1   \\
     1 &   & 0
\end{array}
\right)                 \nonumber \\
&=&\alpha\Sigma_{1}+\beta\Sigma_{1}^{2}. 
\end{eqnarray}
By the way, 
\[
(0,\ \beta,\ \alpha)=(x_{1},\ x_{2},\ x_{3})\ 
\frac{1}{3}
\left(
\begin{array}{ccc}
1&   1&   1              \\
1& \sigma^{2} & \sigma   \\
1& \sigma & \sigma^{2}
\end{array}
\right).
\]
Conversely, 
\[
(x_{1},\ x_{2},\ x_{3})=(0,\ \beta,\ \alpha)
\left\{
\frac{1}{3}
\left(
\begin{array}{ccc}
1&   1&   1              \\
1& \sigma^{2} & \sigma   \\
1& \sigma & \sigma^{2}
\end{array}
\right)
\right\}^{-1}
=
(0,\ \beta,\ \alpha)
\left(
\begin{array}{ccc}
1&   1&   1              \\
1& \sigma & \sigma^{2}   \\
1& \sigma^{2} & \sigma
\end{array}
\right).
\]
Therefore we have 
\begin{equation}
x_{1}=\alpha+\beta,\quad 
x_{2}=\sigma^{2}\alpha+\sigma \beta,\quad 
x_{3}=\sigma \alpha+\sigma^{2}\beta.
\end{equation}
Compare this with (\ref{eq:Cardano formula}). 
Let us solve (\ref{eq:3-degree matrix equation adjoint}). 
\begin{eqnarray}
0&=&(\alpha\Sigma_{1}+\beta\Sigma_{1}^{2})^{3}+
    3p(\alpha\Sigma_{1}+\beta\Sigma_{1}^{2})+qE        \nonumber \\
&=&(\alpha^{3}+\beta^{3}+q)E+
 3\alpha(\alpha\beta+p)\Sigma_{1}+3\beta(\alpha\beta+p)\Sigma_{1}^{2}
\end{eqnarray}
Therefore 
\begin{equation}
\label{eq:alpha-beta}
\left\{
\begin{array}{ll}
   \alpha^{3}+\beta^{3}=-q \\
   \alpha\beta=-p
\end{array}
\right.
\end{equation}
Since $\beta=-p/\alpha$, we have 
\[
\alpha^{3}-\frac{p^{3}}{\alpha^{3}}=-q
\quad \Longrightarrow \quad 
\alpha^{6}+q\alpha^{3}-p^{3}=0.
\]
If we set $t=\alpha^{3}$, then $t^{2}+qt-p^{3}=0$, the solutions are just 
\begin{equation}
t=\frac{-q+\sqrt{q^{2}+4p^{3}}}{2},\quad 
\frac{-q-\sqrt{q^{2}+4p^{3}}}{2}.
\end{equation}
Since $\alpha$ and $\beta$ are symmetric in the equation 
(\ref{eq:alpha-beta}), $\alpha$ and $\beta$ are primitive solutions of 
equations 
\begin{equation}
\alpha^{3}=\frac{-q+\sqrt{q^{2}+4p^{3}}}{2},\quad 
\beta^{3}=\frac{-q-\sqrt{q^{2}+4p^{3}}}{2}.
\end{equation}
This is our derivation of {\bf Cardano formula}.

\section{Ferrari Formula}

In this section we consider an algebraic equation with four degrees. 
By the same reasoning in the preceding section we have only to treat 
\begin{equation}
\label{eq:4-degree equation modify}
x^{4}+px^{2}+qx+r=0
\end{equation}
For this equation we have the famous solutions called {\bf Ferrari formula} : 
Let $\lambda_{0}$ be a solution of the (decomposition) equation with three 
degrees 
\begin{equation}
\lambda^{3}-p\lambda^{2}-4r\lambda+(4pr-q^{2})=0.
\end{equation}
Then four solutions of (\ref{eq:4-degree equation modify}) are given by 
ones of following two equations with two degrees
\begin{equation}
x^{2}\pm\sqrt{\lambda_{0}-p}\left\{x-\frac{q}{2(\lambda_{0}-p)}\right\}
+\frac{\lambda_{0}}{2}=0. 
\end{equation}

\vspace{5mm}
This formula is not similar to that of Cardano. From this formula we cannot 
see the principle of superposition (of solutions), so we are dissatisfied 
at this one. 
From here we reconstruct the solutions by using generalized Pauli and 
Walsh--Hadamard matrices (\ref{eq:sigma-1-four}), (\ref{eq:w-a-four}) in 
four level systems to make the principle of superposition clear. 
Our method may be fresh to not only students (non--experts) 
but also mathematicians. 

Let $x_{1}$, $x_{2}$, $x_{3}$, $x_{4}$ be four solutions of 
(\ref{eq:4-degree equation modify}) and set 
\begin{equation}
A=
\left(
\begin{array}{cccc}
x_{1} &       &       &       \\
      & x_{2} &       &       \\
      &       & x_{3} &       \\
      &       &       & x_{4}
\end{array}
\right),\quad x_{j}^{4}+px_{j}^{2}+qx_{j}+r=0.
\end{equation}
Remind that $x_{1}+x_{2}+x_{3}+x_{4}=0$ from 
(\ref{eq:4-degree equation modify}), so $\mbox{tr}A=0$. Then we have 
\begin{equation}
\label{eq:4-degree matrix equation}
A^{4}+pA^{2}+qA+rE=0,
\end{equation}
where $E$ is a unit matrix. 
Similarly in the preceding section we consider superpositions of solutions 
$x_{j}$ ($j=1, 2, 3, 4$). We operate $W$ in (\ref{eq:w-a-four}) on 
(\ref{eq:4-degree matrix equation}) as the adjoint action ($W^{\dagger}W=E$) 
\begin{equation}
\label{eq:4-degree matrix equation adjoint}
(W^{\dagger}AW)^{4}+p(W^{\dagger}AW)^{2}+qW^{\dagger}AW+rE=0.
\end{equation}
The calculation of $W^{\dagger}AW$ is as follows : 
\begin{eqnarray}
\label{eq:calculation-four}
&&W^{\dagger}AW          \nonumber \\
=&&
\left(
\begin{array}{cccc}
   0       & \beta  & \gamma & \alpha   \\
   \alpha  & 0      & \beta  & \gamma   \\
   \gamma  & \alpha & 0      & \beta    \\
   \beta   & \gamma & \alpha & 0
\end{array}
\right)        
=
\alpha
\left(
\begin{array}{cccc}
   0 &   &    & 1  \\
   1 & 0 &    &    \\
     & 1 & 0  &    \\
     &   & 1  & 0
\end{array}
\right)+
\gamma
\left(
\begin{array}{cccc}
   0 &   & 1  &    \\
     & 0 &    & 1  \\
   1 &   & 0  &    \\
     & 1 &    & 0
\end{array}
\right)+
\beta
\left(
\begin{array}{cccc}
   0 & 1 &    &    \\
     & 0 & 1  &    \\
     &   & 0  & 1  \\
   1 &   &    & 0
\end{array}
\right)         \nonumber \\
=&&\ \alpha\Sigma_{1}+\gamma\Sigma_{1}^{2}+\beta\Sigma_{1}^{3},
\end{eqnarray}
where 
\begin{eqnarray}
0&=&\frac{1}{4}(x_{1}+x_{2}+x_{3}+x_{4}), \quad 
\alpha=\frac{1}{4}(x_{1}+\sigma x_{2}+\sigma^{2}x_{3}+\sigma^{2}x_{4}), 
\nonumber \\
\gamma&=&\frac{1}{4}(x_{1}+\sigma^{2}x_{2}+x_{3}+\sigma^{2}x_{4}), \quad 
\beta=\frac{1}{4}(x_{1}+\sigma^{3} x_{2}+\sigma^{2}x_{3}+\sigma x_{4})
\end{eqnarray}
and $\sigma=\mbox{e}^{\frac{2\pi i}{4}}=i$. Then 
\[
(0,\ \beta,\ \gamma,\ \alpha)=(x_{1},\ x_{2},\ x_{3},\ x_{4})\ 
\frac{1}{4}
\left(
\begin{array}{cccc}
1 & 1          &   1        & 1           \\
1 & \sigma^{3} & \sigma^{2} & \sigma      \\
1 & \sigma^{2} & 1          & \sigma^{2}  \\
1 & \sigma     & \sigma^{2} & \sigma^{3} 
\end{array}
\right).
\]
Conversely, 
\begin{eqnarray}
&&(x_{1},\ x_{2},\ x_{3},\ x_{4})         \nonumber \\
=
&&(0,\ \beta,\ \gamma,\ \alpha)
\left\{
\frac{1}{4}
\left(
\begin{array}{cccc}
1 & 1          &   1        & 1           \\
1 & \sigma^{3} & \sigma^{2} & \sigma      \\
1 & \sigma^{2} & 1          & \sigma^{2}  \\
1 & \sigma     & \sigma^{2} & \sigma^{3} 
\end{array}
\right)
\right\}^{-1}   
=
(0,\ \beta,\ \gamma,\ \alpha)
\left(
\begin{array}{cccc}
1 & 1          &   1        & 1           \\
1 & \sigma     & \sigma^{2} & \sigma^{3}  \\
1 & \sigma^{2} & 1          & \sigma^{2}  \\
1 & \sigma^{3} & \sigma^{2} & \sigma 
\end{array}
\right),              \nonumber 
\end{eqnarray}
so we have 
\begin{eqnarray}
x_{1}&=&\alpha+\gamma+\beta,\quad 
x_{2}=\sigma^{3}\alpha+\sigma^{2}\gamma+\sigma \beta,      \nonumber \\
x_{3}&=&\sigma^{2}\alpha+\gamma+\sigma^{2}\beta,\quad 
x_{4}=\sigma \alpha+\sigma^{2}\gamma+\sigma^{3}\beta. 
\end{eqnarray}
This is just superpositions of $\alpha$, $\gamma$ and $\beta$. 
Next let us determine them. From (\ref{eq:calculation-four}) 
\[
(W^{\dagger}AW)^{2}=
(\gamma^{2}+2\alpha\beta)E+
2\gamma\beta\Sigma_{1}+
(\alpha^{2}+\beta^{2})\Sigma_{1}^{2}+
2\alpha\gamma\Sigma_{1}^{3}
\]
and 
\begin{eqnarray}
(W^{\dagger}AW)^{4}=\{(W^{\dagger}AW)^{2}\}^{2}=
&&
\left\{
(\alpha^{2}+\beta^{2})^{2}+(\gamma^{2}+2\alpha\beta)^{2}+
8\alpha\beta\gamma^{2}
\right\}E+              \nonumber \\
&&
4\left\{
(\alpha^{2}+\beta^{2})\alpha+(\gamma^{2}+2\alpha\beta)\beta
\right\}\gamma\Sigma_{1} +   \nonumber \\
&&
2(\alpha^{2}+\beta^{2})(3\gamma^{2}+2\alpha\beta)\Sigma_{1}^{2}+  \nonumber \\
&&
4\left\{
(\alpha^{2}+\beta^{2})\beta+(\gamma^{2}+2\alpha\beta)\alpha
\right\}\gamma\Sigma_{1}^{3}.     \nonumber 
\end{eqnarray}
Then the equation (\ref{eq:4-degree matrix equation adjoint}) gives 
\begin{equation}
\label{eq:alpha-beta-gamma}
\left\{
\begin{array}{ll}
 (\alpha^{2}+\beta^{2})^{2}+(\gamma^{2}+2\alpha\beta)^{2}+
 8\alpha\beta\gamma^{2}+p(\gamma^{2}+2\alpha\beta)+r=0        \\
 4\left\{
(\alpha^{2}+\beta^{2})\alpha+(\gamma^{2}+2\alpha\beta)\beta
\right\}\gamma+2p\gamma\beta+q\alpha=0                        \\
2(\alpha^{2}+\beta^{2})(3\gamma^{2}+2\alpha\beta)+p(\alpha^{2}+\beta^{2})
+q\gamma=0                                                    \\
4\left\{
(\alpha^{2}+\beta^{2})\beta+(\gamma^{2}+2\alpha\beta)\alpha
\right\}\gamma+2p\alpha\gamma+q\beta=0 
\end{array}
\right.
\end{equation}
Now let us transform the equations into a more convenient form 
(this point is important). 
\begin{equation}
\label{eq:alpha-beta-gamma-transform}
\left\{
\begin{array}{ll}
(a)\quad (\alpha^{2}+\beta^{2})^{2}+(\gamma^{2}+2\alpha\beta)^{2}+
 8\alpha\beta\gamma^{2}+p(\gamma^{2}+2\alpha\beta)+r=0        \\
(b)\quad 2\gamma\beta(2\gamma^{2}+6\alpha\beta+p)+4\alpha^{3}\gamma+
q\alpha=0                                                     \\
(c)\quad (\alpha^{2}+\beta^{2})(6\gamma^{2}+4\alpha\beta+p)+
q\gamma=0                                                    \\
(d)\quad 2\alpha\gamma(2\gamma^{2}+6\alpha\beta+p)+4\beta^{3}\gamma+
q\beta=0                                                     
\end{array}
\right.
\end{equation}
From (b) and (d) 
\[
 -\frac{4\alpha^{3}\gamma+q\alpha}{2\gamma\beta}
=2\gamma^{2}+6\alpha\beta+p
=-\frac{4\beta^{3}\gamma+q\beta}{2\alpha\gamma},
\]
so 
\[
4\alpha^{4}\gamma+q\alpha^{2}=4\beta^{4}\gamma+q\beta^{2}
\Longrightarrow 
(\alpha^{2}-\beta^{2})\left\{4(\alpha^{2}+\beta^{2})\gamma+q\right\}=0,
\]
so we have 
\begin{equation}
\label{eq:sub-equation-1}
\alpha^{2}+\beta^{2}=-\frac{q}{4\gamma}.
\end{equation}
Substituting this into (c) 
\begin{equation}
\label{eq:sub-equation-2}
6\gamma^{2}+4\alpha\beta+p-4\gamma^{2}=0
\Longrightarrow 
\gamma^{2}+2\alpha\beta=-\frac{p}{2}\quad \mbox{or}\quad 
2\alpha\beta=-\gamma^{2}-\frac{p}{2}.
\end{equation}
Substituting these into (a) 
\[
\left(-\frac{q}{4\gamma}\right)^{2}+
\left(-\frac{p}{2}\right)^{2}
-4\left(\gamma^{2}+\frac{p}{2}\right)\gamma^{2}
-\frac{p^{2}}{2}+r=0.
\]
Rearranging this equation 
\[
64\gamma^{6}+32p\gamma^{4}-4(4r-p^{2})\gamma^{2}-q^{2}=0.
\]
If we set $\Gamma=\gamma^{2}$, then wa obtain the algebraic equation with 
three degrees 
\begin{equation}
\label{eq:reduction-equation}
64\Gamma^{3}+32p\Gamma^{2}-4(4r-p^{2})\Gamma-q^{2}=0.
\end{equation}
Since this equation can be solved (in the preceding section), we obtain 
$\gamma=\sqrt{\Gamma}$ for the largest real solution $\Gamma$. 
Next we look for $\alpha$ and $\beta$. 
From (\ref{eq:sub-equation-1}) and (\ref{eq:sub-equation-2}) 
\begin{equation}
\left\{
\begin{array}{ll}
\alpha^{2}+\beta^{2}=-\frac{q}{4\gamma}       \\
\alpha\beta=-\frac{1}{2}(\Gamma+\frac{p}{2})
\end{array}
\right.
\end{equation}
By the same method in (\ref{eq:alpha-beta}) $\alpha$ and $\beta$ are 
primitive solutions of equations 
\begin{eqnarray}
\alpha^{2}&=&\frac{1}{2}
\left\{
-\frac{q}{4\gamma}+
\sqrt{\left(\frac{q}{4\gamma}\right)^{2}
-\left(\Gamma+\frac{p}{2}\right)^{2}}
\right\},                                    \\
\beta^{2}&=&\frac{1}{2}
\left\{
-\frac{q}{4\gamma}-
\sqrt{\left(\frac{q}{4\gamma}\right)^{2}
-\left(\Gamma+\frac{p}{2}\right)^{2}}
\right\}. 
\end{eqnarray}
Namely, we determined $\alpha$, $\beta$ and $\gamma$ completely. 

\par \noindent
This is our derivation (a variation) of {\bf Ferrari formula}.

\section{A (Real) Problem}

In this section let us consider an algebraic equation with five degrees. 
\begin{equation}
\label{eq:5-degree equation modify}
x^{5}+px^{3}+qx^{2}+rx+s=0.
\end{equation}
We can apply the same method in the case of three and four degrees to 
this equation, however we cannot solve this one in an algebraic manner 
as in the preceding sections (Abel and Galois). 
We know the Galois theory to decide whether an algebraic equation has 
algebraic solutions or not. This is a classical theory. 

Then we have some natural questions : 
Is there some relation between the Galois theory and the principle of 
superposition ? 
What is a role of Galois theory in general Qudit theory ? 
Moreover, is it possible to consider a quantum Galois theory in general 
Qudit theory ? 

These are interesting problems in Quantum Computation.

\vspace{10mm}
\noindent{\em Acknowledgment.}\\
The author wishes to thank Shin'ichi Nojiri for his helpful comments and 
suggestions.

\vspace{10mm}
\begin{center}
\begin{Large}
\noindent{\bfseries Appendix\quad Euler Formula}
\end{Large}
\end{center}
\vspace{5mm}

Akira Asada pointed out to me that my method to derive the variation of 
Ferrari formula was similar to that of Euler which I didn't know when 
writing this note. In fact his point was correct, so let us copy the 
Euler formula in \cite{IY}
\footnote{
I could not find a textbook written in English which comments on the formula 
}
for the completeness (this formula is not necessarily popular even in 
Mathematics). 

For the quartic equation (\ref{eq:4-degree equation modify}) we consider the 
cubic equation 
\begin{equation}
t^{3}+\frac{p}{2}t^{2}+\left(\frac{p^{2}}{16}-\frac{r}{4}\right)t-
\frac{q^{2}}{64}=0.
\end{equation}
We note that this equation is just our (reduction) equation 
(\ref{eq:reduction-equation}). Since this equation has three solutions 
($t_{1}$, $t_{2}$, $t_{3}$), we choose signs of square roots 
$\sqrt{t_{1}}$, $\sqrt{t_{2}}$, $\sqrt{t_{3}}$ to satisfy
\begin{equation}
\sqrt{t_{1}}\sqrt{t_{2}}\sqrt{t_{3}}=-\frac{q}{8}.
\end{equation}
Then the solutions are given by 
\begin{eqnarray}
x_{1}&=& \sqrt{t_{1}}+\sqrt{t_{2}}+\sqrt{t_{3}}, \quad 
x_{2}= \sqrt{t_{1}}-\sqrt{t_{2}}-\sqrt{t_{3}},         \nonumber \\
x_{3}&=&-\sqrt{t_{1}}+\sqrt{t_{2}}-\sqrt{t_{3}}, \quad 
x_{4}=-\sqrt{t_{1}}-\sqrt{t_{2}}+\sqrt{t_{3}}.
\end{eqnarray}
To derive the solutions one use various relations between solutions (roots) 
and coefficients, which is quite complicated. 

\vspace{5mm}
Our method is just a modern derivation of the Cardano and Euler formulas 
which doesn't use relations between solutions (roots) and coefficients 
in an apparent manner.


\end{document}